\documentclass[seceq,preprint,eqsecnum,showpacs,preprintnumbers,amsmath,amssymb,floatfix]{jpsj2}

\def\runtitle{Carrier States and Ferromagnetism 
in Diluted Magnetic Semiconductors }
\def\runauthor{Masao \textsc{Takahashi} and Kenn \textsc{Kubo}}
\title{%
Carrier States and Ferromagnetism in Diluted Magnetic Semiconductors}
\author{
Masao \textsc{Takahashi} 
\thanks{E-mail address: taka@gen.kanagawa-it.ac.jp}
and Kenn \textsc{Kubo}$^{1}$}
\inst{%
Kanagawa Institute of Technology \\
1030 Shimo-Ogino, Atsugi-shi, 243-0292 \\
$^1$Department of Physics,  
Aoyama Gakuin University \\
Sagamihara, Kanagawa   229-8558
}
\recdate{\today}

\abst{%
 Applying the dynamical coherent potential approximation to a simple model,
we have systematically studied the carrier states
 in $A_{1-x}$Mn$_xB$-type diluted magnetic semiconductors (DMS's).
The model calculation was performed for three typical cases of DMS's:
The cases with  strong and moderate exchange interactions in the absence
of 
nonmagnetic potentials,  and the case with  strong attractive nonmagnetic
potentials in addition to moderate exchange interaction.
When the exchange interaction is sufficiently strong, 
magnetic impurity bands split from the host band.
Carriers in the magnetic impurity band mainly stay at magnetic sites,
and coupling between the carrier spin and the localized spin is very strong.
The hopping of the carriers among the magnetic sites causes ferromagnetism
through a {\it double-exchange (DE)-like} mechanism.
We have investigated the condition for the DE-like mechanism to operate
in DMS's.
The result reveals that the nonmagnetic attractive potential at the
magnetic site 
assists the formation of the magnetic impurity band
and makes the DE-like mechanism operative by substantially
 enhancing the effect of the exchange interaction.
Using conventional parameters
we have studied the carrier states in Ga$_{1-x}$Mn$_x$As.
The result shows that the ferromagnetism is caused 
through the DE-like mechanism by the carriers in the 
bandtail originating from the impurity states.
}
\kword{%
diluted magnetic semiconductors, exchange interaction, coherent
potential approximation,
carrier-induced ferromagnetism
}
\begin{document}
\sloppy
\maketitle

\section{Introduction}
For more than two decades
diluted magnetic semiconductors (DMS's) have attracted much attention
because of the combination of magnetic and semiconducting properties.
In $A^{\rm II}_{1-x}$Mn$_xB^{\rm VI}$-type (II-VI-based) DMS's, 
Mn impurities substituting for 2+ cations act as stable 2+ ions
and therefore there are few carriers, making these DMS's insulators. 
It is widely accepted that in II-VI-based DMS's 
 a carrier (\textit{s} electron or \textit{p} hole) 
moves over many sites while interacting with the localized (\textit{d})
spins on Mn sites
through the \textit{sp-d} exchange interaction \cite{Furdyna88}.
The exchange interaction strongly enhances
 the effect of magnetic field on band splitting,
leading to spectacular magnetooptical effects
 (e.g., giant Faraday rotation or Zeeman splitting).
In recent years the attention has also been focused on III-V-based DMS's
(Ga$_{1-x}$Mn$_x$As and In$_{1-x}$Mn$_x$As)
due to the high potential for new device applications.
It is highly noteworthy that doping of Mn into GaAs and InAs leads to ferromagnetism,
and interesting magnetooptical and magnetotransport phenomena.
This ferromagnetism is generally called ``carrier-induced ferromagnetism''
because hole carriers introduced by Mn incorporation mediate 
the ferromagnetic coupling between Mn ions \cite{Ohno99}.
The microscopic mechanism for carrier-induced ferromagnetism is still controversial.
The following properties, however, seem to be generally accepted as for (Ga,Mn)As:
(i) Mn ions substitute randomly for Ga cations in the zincblende
structure \cite{Ohno99}.
(ii) A Mn ion in GaAs gives rise to an acceptor level at about 0.113 eV
above the valence band
\cite{Linnarsson97}. 
(iii) The Mn ion has highly localized \textit{d} states with a magnetic moment
of $\sim 5\mu _B$ (or $S=5/2$) \cite{Linnarsson97,Iye99,Ohldag00}.
(iv) The Mn-induced states near the Fermi energy play a key role in the
origin of ferromagnetism.
According to photoemission studies \cite{Oka98,Oka99,Oka01},
 X-ray absorption spectroscopy \cite{Ishiwata02},
 and band calculations \cite{Shirai98,Park01},
those states are mainly created in As 4\textit{p} orbits.
(v) The \textit{p-d} exchange interaction between the As 4\textit{p}
hole 
and the localized \textit{d} spin is antiferromagnetic \cite{Ando98,Szczytko99},
and its amplitude is not very different from that in
II-VI-based DMS's \cite{Oka98,Oka02}. 
(vi) Antisite defects (As ions sitting on Ga lattice site),
 for example, are common in semiconductor samples
grown by low-temperature molecular beam epitaxy\cite{Hayashi01}.  
 Many holes may be trapped not at Mn acceptors but
at such defects, though we may expect one hole donated by a Mn atom.
The density of the holes and that of the Mn ions are therefore regarded
as separate sample-dependent quantities
that are to be determined experimentally.
 
Since the fabrication of III-V DMS's much effort has been made on the
elucidation of the origin of the ferromagnetism. 
To date, 
 the Ruderman-Kittel-Kasuya-Yosida (RKKY) mechanism \cite{Matsukura98},
 the double-exchange mechanism based on $d$ electron hopping
\cite{Akai98}, 
the spin-wave approximation for the kinetic-exchange model \cite{Konig00},
the double-resonance mechanism \cite{Inoue00}, the 
Zener-model description \cite{Dietl00}, 
a theory based on the random-phase approximation and the
coherent potential approximation \cite{Bouzerar02},
 and
the coherent potential approach to a model with 
randomly distributed Ising spins \cite{Kayanuma02}
 have been proposed.
To our knowledge, however,  
  no systematic theoretical study on the carrier states in DMS's, in which
 both the exchange interactions and nonmagnetic attractive potentials
 exist,  have been accomplished  taking the  effect of
 disorder into account. 
 
In this study we aim to clarify the nature of carrier states in DMS's 
applying the dynamical coherent potential approximation 
(dynamical CPA) to a simple model. 
Furthermore, on the basis of the numerical calculation, we discuss the
origin and the mechanism of the carrier-induced ferromagnetism in 
III-V DMS's \cite{taka02}.

The main content and the organization of this paper are as follows.
In \S 2, we briefly summarize the present model and the dynamical CPA.
In \S 3 after a general consideration of the DMS's and the model parameters
 (\S 3.1),
we first present the results for the carrier states in 
the DMS's with no nonmagnetic attractive potential:
the case of the strong exchange interaction  (\S 3.2)
and the case of the moderate exchange interaction (\S 3.3).
Then, based on the result 
we discuss the mechanism of carrier-induced ferromagnetism in DMS's (\S 3.4).
The result of this study reveals that the double-exchange (DE)-like
mechanism \cite{Zener51,Anderson55}
is in effect to realize a high Curie temperature
 when the exchange interaction is strong enough to form
a magnetic impurity band split from the host band.
On the other hand, the RKKY mechanism 
may be relevant when the exchange interaction is so weak as not to produce
 a magnetic impurity band.
The latter corresponds to the case of II-VI-based DMS's.
As mentioned above, the strength of the \textit{p-d} exchange
interaction in
 III-V DMS's  
is not very different from that in II-VI DMS's. 
Then, what causes the ferromagnetism in III-V DMS ?
The key lies in the attractive Coulomb potential
between a \textit{p} hole and a Mn acceptor center. 
In \S 3.5, we investigate the role of the attractive potential
to elucidate the origin of the carrier-induced ferromagnetism in III-V DMS's.
Section 4 is devoted to (Ga,Mn)As together with a comparison with (In,Mn)As.
Concluding remarks are presented in \S 5.

\section{Basic consideration}
\subsection{Model Hamiltonian for a carrier in a DMS}
In order to study the effect of the \textit{p-d} exchange interaction 
between a carrier (\textit{p} hole) 
and localized magnetic moments together with
 magnetic and chemical disorder in DMS's,
 we previously proposed a simple model \cite{taka99,taka02}.
In the model, the local potentials of nonmagnetic $A$ ions
 in a semiconducting compound $AB$ are 
substituted randomly, with mole fraction $x$, 
by the local potentials that include the exchange interactions 
between  carrier spins and the localized spin moments on  magnetic
\textit{M} ions.
 Thus, the potential to which a carrier is subject at a site differs
depending on 
whether the site is occupied by an \textit{A} ion or an \textit{M} ion.
The Hamiltonian $H$ is given by
\begin{eqnarray}
\label{Hamiltonian}
   H  & = &  \sum_{m,n,\mu} \varepsilon_{mn}  a^{\dag} _{m\mu} a_{n\mu}
\  
          + \sum_{n} u_n  , 
\end{eqnarray}
where $u_n$ is
 either $u_n^A$ or $u_n^M$, depending on the ion species occupying the
\(n\) site:
\begin{eqnarray}
   u_n^A &=& E_A  \sum_{\mu}
 a^{\dag} _{n\mu} a_{n\mu} \  , \\
   u_n^M &=& E_M  \sum_{\mu} 
a^{\dag}_{n\mu} a_{n\mu}  
   - I \sum_{\mu,\nu} a^{\dag}_{n\mu} \mbox{\boldmath $\sigma$}_{\mu \nu }
    \cdot {\bf S}_{n} a_{n\nu} \ .
\end{eqnarray}
The notations here are conventional and the same as in the previous papers
\cite{taka99,taka01a,taka01b,taka02}. 
Here,
$E_A$ ($E_M$) represents a nonmagnetic local potential at  an $A$ ($M$) site.
We employ a local potential $E_M$
although 
the real potential for
a hole is an attractive screened
Coulomb potential exerted  by a  Mn$^{2+}$  acceptor center
in III-V-based DMS's such as Ga$_{1-x}$Mn$_x$As. 
 The exchange interaction between a carrier spin and the localized spin 
 $\textbf{S}_n$ of the Mn site \textit{n}
is expressed by 
$-I  a^{\dag}_{n\mu} \mbox{\boldmath $\sigma$}_{\mu \nu } \cdot {\bf
S}_n  a_{n\nu} $,
where $ \mbox{\boldmath $\sigma$}_{\mu \nu }$ represents the element of
 the Pauli spin matrices. 

\subsection{Dynamical coherent potential approach}
 A carrier moving in a DMS described by Eq.\ (\ref{Hamiltonian})
 is subject to disordered potentials
which arise not only from substitutional disorder 
but also from thermal fluctuations of \textit{d} spins
through the \textit{p-d} exchange interaction. 
Furthermore, when magnetization arises,
the effective potential for the carrier differs 
 according to the orientation of the carrier's spin.
In the dynamical coherent potential approximation (dynamical CPA)
\cite{Kubo74,taka96}, 
the disordered potential is taken into account in terms of the
spin-dependent effective medium
 where a carrier is subject to a coherent potential
 $\Sigma_{\uparrow}$ or $\Sigma_{\downarrow}$
according to the orientation of its spin.
The coherent potential $\Sigma_{\uparrow}$ ($\Sigma_{\downarrow}$)
 is an energy ($\omega$) 
dependent complex potential.
 Then, a carrier moving in this effective medium 
 is described by the unperturbed Hamiltonian $K$:
\begin{eqnarray}
    K & = &  \sum_{k\mu}(\varepsilon_k + \mit\Sigma_{\mu})
                a\sp{\dag} _{k\mu} a_{k\mu}  \    .
\end{eqnarray}
Thus, the perturbation term $V (= H - K)$ is written 
 as a sum over each lattice site: 
\begin{eqnarray}
   V  & = &   \sum_{n}v_n \  ,   \ 
\end{eqnarray}
where $v_n$ is
 either $v_n^A$ or $v_n^M$, depending on the ion species occupying the
$n$ site:
\begin{eqnarray}
   v_n^A &=& \sum_{\mu} (E_A-\Sigma _{\mu }) a^{\dag} _{n\mu} a_{n\mu} 
\  , \\
   v_n^M &=& \sum_{\mu}( E_M -\Sigma _{\mu }) a^{\dag} _{n\mu} a_{n\mu} 
\  
   - I \sum_{\mu,\nu} a^{\dag}_{n\mu} \mbox{\boldmath $\sigma$}_{\mu \nu }
    \cdot {\bf S}_{n} a_{n\nu} \ .
\end{eqnarray}
Next, using the reference Green's function $G_0$ given by
\begin{eqnarray}
  \langle m\mu\ | G_0(\omega)\ | n\nu\rangle  & = &
 \left\langle a_{m\mu}\frac{1}{\omega - K } a ^{\dag}
_{n\nu}\right\rangle_0   ,  \ 
\end{eqnarray}
 where $\langle O \rangle_0$ denotes 
the expectation value of $O$ in the vacuum, 
we define the $t$-matrix $t^A$ which represents the multiple scattering of
carriers due to an $A$ ion 
embedded in the effective medium by
\begin{eqnarray}
   t^A_n & = & v^A_n  [1-G_0 v^A_n]^{-1}      \ , 
\label{tmatrixA}
\end{eqnarray}
and  $t^M$  due to an $M$ ion  by
\begin{eqnarray}
   t^M_n & = & v^M_n  [1-G_0 v^M_n]^{-1}      \ . 
\label{tmatixM}
\end{eqnarray}
We have omitted suffices of the matrices $G$, $t$ and $v$ since they are 
self-evident. 
The matrix $t^A_n$ ($t^M_n$) represents the  scattering 
 by an isolated potential $v^A_n$ ($v^M_n$) in the effective medium completely.
Note that $K$, and thus $G_0$, includes no localized spin operator.
 According to the multiple-scattering theory,
the total scattering operator $T$ of a random medium, which is related
 to
 ${\displaystyle
  G \equiv \left\langle a\frac{1}{\omega -H}a^{\dag} \right\rangle_0 }$ as
\begin{eqnarray}
   G & = & G_0 + G_0 T G_0      \  ,
\end{eqnarray}
is expressed as a series,
\begin{eqnarray}
   T & = & \sum_n t_n + \sum_n t_n G_0 \sum_{m \ ( \neq n ) } t_m  
 + \sum_n t_n G_0 \sum_{m \  ( \neq n ) } t_m  G_0 \sum_{l\  ( \neq m )
} t_l + \cdots    \  .
\end{eqnarray}
In the CPA,  the condition 
\begin{eqnarray}
   \langle t_n \rangle_{\rm av} & = &  0  \qquad \mbox{ for any site
{\it n} }     \  
\end{eqnarray}
determines the coherent potential $\Sigma_{\mu} (\mu = \uparrow ,
\downarrow )$
 and we approximate 
 $\langle G \rangle_{\rm av} $ by $G_0$.
Here we express the average of a quantity $O$ over the disorder in the system
as $ \langle O \rangle_{\rm av}$.
Since the present system includes both substitutional disorder
and the thermal fluctuations of the localized spin an $M$ site,
the average of the $t$ matrix is written as 
\begin{eqnarray}
   \langle t_n \rangle_{\rm av} & = &
  (1-x) t^A_n + x \langle t^M_n (\textbf{S})  \rangle   \ .
\end{eqnarray}
Here $(1-x)$ and \textit{x} are the mole fractions of \textit{A} and
\textit{M} atoms, respectively;
 $\langle t^M(\textbf{S}) \rangle $ means
 the thermal average of $t^M$ over fluctuating localized spin
\textbf{S}. 
 The coherent potential $ \Sigma _{\mu }$
is decided such that the effective scattering of a carrier at the chosen
site 
embedded in the effective medium is zero on average.
Since the spin off-diagonal element becomes zero after the average, 
the condition (the dynamical CPA condition) is given by
\begin{subequations}
\begin{eqnarray}
(1-x) t^A_{\uparrow \uparrow} + x \langle t^M_{\uparrow \uparrow}(S_z)
\rangle &=& 0 \ , 
\label{CPAup}  \\
(1-x) t^A_{\downarrow \downarrow} + x \langle t^M_{\downarrow
\downarrow}(S_z) \rangle &=& 0  \ .
\label{CPAdown}
\end{eqnarray}
\label{CPAcondition}
\end{subequations}
For simplicity, the \textit{t} matrix elements 
 in the site representation  
 $<n \mu|t|n \nu >$ ($n$ is a site index, $\mu, \nu  = \uparrow$ or $ \downarrow$)
are written as $t_{\mu \nu}$.
It is worth noting that in the calculation of $t^M_{\uparrow \uparrow} 
\ (t^M_{\downarrow \downarrow}$) the spin flip processes are  properly
taken into account, and  a single \textit{t}-matrix element
 $t^M_{\mu \mu }$ depends
on both $\Sigma _{\uparrow }$ and $\Sigma _{\downarrow }$.
Therefore we need to solve Eqs. (\ref{CPAup}) and (\ref{CPAdown}) simultaneously.
Note that the diagonal matrix element $t^M_{\mu \mu }$ 
 involves $S_z$ as an operator,
where $S_z$ stands for the $z$-component of the localized spin. 
The thermal average over the fluctuating localized spin 
 is taken as 
\begin{equation}
\langle t^M_{\mu \mu}(S_{z})\rangle
 = \sum_{S_z=-S}^S t^M_{\mu \mu}(S_z)
{\rm exp} \left(\frac{hS_z}{k_BT} \right)
/\sum_{S_z=-S}^S {\rm exp} \left(\frac{hS_z}{k_BT}\right) \ ,
\label{THREM}
\end{equation}
where $h$ denotes the effective field felt by the localized spins. 
Since there is one-to-one correspondence between
 $\langle S_z\rangle$ and the parameter
 $\lambda \equiv h/k_BT$, 
we can describe the carrier states in terms of 
$\langle S_z\rangle$ instead of $\lambda$.
Note that 
the thermal average of off-diagonal $t$-matrix elements 
 $\langle t^M_{\uparrow \downarrow } \rangle_{\rm av}
 = \langle t^M_{\downarrow \uparrow } \rangle_{\rm av} =0$
 because the magnetization is assumed to be along the $z$-axis.
In this work we treat the localized 
 spins classically for simplicity. 
The actual calculations were performed for $S=400$.

Throughout this work we assume a model density of states 
with  a semicircular profile whose  half-bandwidth is $\Delta$, 
\begin{eqnarray}
 \rho (\varepsilon) 
  & = & \frac{2}{\pi\Delta} 
            \sqrt{1- \left(\frac{\varepsilon}{\Delta}\right)^2 } \ ,
\label{rho}
\end{eqnarray}
as the unperturbed density of states. 
Then the density of states with $\mu$ spin $D_{\mu}(\omega)$ is
calculated as
\begin{eqnarray} 
D_{\mu}(\omega) &=&  -\frac{1}{\pi} {\rm Im} \langle n\mu |G(\omega
)|n\mu  \rangle_{\rm av}
  \cong  -\frac{1}{\pi} {\rm Im} \langle n \mu |G_0(\omega ) |n\mu 
\rangle \nonumber \\
 &=&  -\frac{1}{\pi} {\rm Im} \int^{\Delta }_{-\Delta }
d\varepsilon \frac{\rho (\varepsilon  )}{\omega -\varepsilon -\Sigma
_{\mu }(\omega )} \ 
\label{Dens}
\end{eqnarray}
by using  $\Sigma _{\mu }$ determined by the  condition (\ref{CPAcondition}).
In all of the present results we  confirmed numerically the sum rule:  
\begin{eqnarray}
\int ^{\infty}_{-\infty} D_{\uparrow }(\omega ) d \omega 
=  \int ^{\infty}_{-\infty} D_{\downarrow }(\omega ) d \omega 
= 1 \ .
\end{eqnarray} 

\subsection{The Curie temperature}
We calculate the Curie temperature $T_c$ to investigate the condition
for the occurrence of ferromagnetism.
Throughout this paper we assume that the carriers are degenerate. 
Then we obtain the  density of the carrier with $\mu $ spin $n_\mu$  
and the total energy $E(\langle S_z\rangle)$  as
\begin{eqnarray}
n_{\mu } &=& \int_{-\infty}^{\varepsilon_F}  
D_{\mu }(\omega) 
d \omega \  
\end{eqnarray}
and 
\begin{eqnarray}
E(\langle S_z\rangle) &=& \int_{-\infty}^{\varepsilon_F}  \omega
[D_{\uparrow}(\omega)+D_{\downarrow}(\omega)] 
d \omega \  ,
\end{eqnarray}
respectively, 
as  functions of the Fermi level $\varepsilon _F$.
Here the dependence on  $\langle S_z\rangle$ is contained
 in $D_{\mu }(\omega)$.
Note  that  $E(\langle S_z\rangle)$ is the sum of the kinetic and the
exchange energies.
 For a fixed value of $\langle S_z\rangle/S$, the total carrier density
$n\ (\equiv n_{\uparrow
}+n_{\downarrow })$
has one-to-one correspondence with  $\varepsilon _F$ and therefore 
  $E(\langle S_z\rangle)$ 
can be expressed  as  a function of $n$.
The free energy per site of the system   is given as
\begin{eqnarray}
F(\langle S_z\rangle) &=& E(\langle S_z\rangle)
 -  T {\cal S} ,
\end{eqnarray}
where the entropy  due to the localized spins is given by
\begin{eqnarray}
 {\cal S} & = & x k_B \log \sum_{S_z = -S}^S {\rm exp} \left( 
\lambda  S_z \right)  - x k_B \lambda  \langle S_z\rangle \ .
\end{eqnarray}
The parameter $\lambda$
is determined so as to minimize $F(\langle S_z\rangle)$ through the
condition 
$\frac{d}{d \lambda } F(\langle S_z\rangle) =0$.
 If we expand   $F(\langle S_z\rangle)$
in terms of $(\langle S_z\rangle)^2$,  $T_c$ is determined as the
temperature 
where the coefficient of $(\langle S_z\rangle)^2$ vanishes. 
As is revealed later [see Fig. 20(a)]
the energy difference  between the ferromagnetic and
 paramagnetic states,

$E(\langle S_z\rangle)-E(0)$, is  
approximately proportional
 to $(\langle S_z\rangle)^2$ up to full polarization at a fixed carrier density.
 Therefore we fitted  the  numerical data of $[E(\langle
S_z\rangle)-E(0)]/\Delta$   to the expansion
\begin{eqnarray}
\frac{E(\langle S_z\rangle)-E(0)}{\Delta} &=& -a \left( \frac{\langle S_z\rangle}{S}\right)^2
+b\left(\frac{\langle S_z\rangle}{S}\right)^4 
\label{Expansion}
\end{eqnarray}
and estimated the coefficients  $a$ and $b$. 
Finally  we obtain $T_c$ as 
\begin{eqnarray}
 \frac{k_B T_c}{\Delta } & = & \frac{2a}{3x} \ .
\label{Tc}
\end{eqnarray}
%

\section{Carrier states in DMS's}
\subsection{General consideration}
Let us start with confirming that the exchange interaction term, 
$-I \mbox{\boldmath $\sigma$} \cdot {\bf S} $,
has two energy eigenvalues
according to the manner of coupling between the carrier spin and the
localized spin.
The parallel coupling state (denoted by p) with $2S+2$-fold degeneracy
has the energy eigenvalue  
$\varepsilon _\textrm{p} = -IS$,
while the antiparallel coupling state (denoted by a)  with $2S$-fold
degeneracy has the energy
eigenvalue 
$\varepsilon _\textrm{a}= I(S+1)$. 
The classical spin approximation  assumes $S$ to be infinite 
while keeping $IS$ constant.
Then  the  parallel and antiparallel eigenstates 
have the same degeneracy  and their energy  eigenvalues  have the same 
absolute 
value $|IS|$. 
In this work we adopt the classical spin approximation since the
magnitude of the localized spin on a  Mn$^{2+}$
($S=\frac{5}{2}$)  is pretty large. 
It is, therefore,  sufficient to appoint the value of the exchange
energy $IS$
instead of appointing the values of $I$ and $S$ separately.

In most DMS's, 
the \textit{p-d} exchange interaction
between a  \textit{p} hole and a localized spin 
favors antiparallel coupling. 
The \textit{p-d} exchange interaction is known to play a crucial role
in magnetooptical effects in II-VI-DMS's,   
and  is also believed to cause  the carrier-induced ferromagnetism 
in III-V-DMS's. Hence, keeping the \textit{p-d} exchange interaction in mind,
we assume $IS <0$ hereafter.
  The present model is characterized by only two parameters,
 $IS/\Delta $ and $E_M/\Delta $, since we take $E_A $ as the origin
($=0$) of the energy.

\subsection{Result for $IS=-\Delta $ and $E_M=0$ }
In the following  we show the results for three typical cases of 
$A_{1-x}M_xB$-type DMS's,
 in which 5 \% of the nonmagnetic ($A$) ions are randomly substituted
by magnetic (\textit{M}) ions.
The present  and next subsections reports typical cases in $E_M=0$.
In Figs. 1 $\sim$ 4 
we show the numerical results for 
 $IS=-\Delta $ and $E_M =0.0$,
which is the case where the exchange interaction is so strong
that magnetic impurity bands appear.
In Fig.\ 1, the spin-polarized DOSs,
 $D_{\uparrow }(\omega )$ and $D_{\downarrow }(\omega )$, 
are depicted for various values of $\langle S_z \rangle /S$.
In the dilute  limit ($x \rightarrow 0$)
impurity levels appear at the energies of 
$\frac{E_a}{\Delta } = \left( \frac{E_M \mp IS}{\Delta }\right)
 + \frac{1}{4} \left(\frac{\Delta}{E_M \mp IS} \right)=\pm 1.25$.
When $x=0.05$, impurity bands form around the impurity levels.
The total number of states of each impurity band is \textit{x},
irrespectively of $\langle S_z \rangle$.
The low (high) energy impurity band corresponds to the antiparallel (parallel)
coupling state.
The impurity bands are strongly affected by the change in $\langle S_z \rangle$.
On the other hand, the host band 
is negligibly affected.
In order to elucidate the origin of the change in the DOS,
we calculate the species-resolved DOS.
In Fig.\ 2, we depict the \textit{A-} and \textit{M}-site components of
the DOS, 
$(1-x)D^A_{\mu }(\omega )$ and $xD^M_{\mu }(\omega )$, 
where $D^A_{\mu }(\omega )$ [$D^M_{\mu }(\omega )$] represents the local
DOS with $\mu $ spin
 ($\mu =\uparrow $ or $\downarrow $) associated with the \textit{A}
 (\textit{M}) ion. 
%
The total number of $A$- and \textit{M}-site states are
$1-x$ and \textit{x}, respectively. Since $D^A_{\mu }(\omega )$ and
$D^M_{\mu }(\omega )$ are
normalized 
\begin{eqnarray}
D_{\mu }(\omega )  = (1-x) D^A_{\mu }(\omega ) + x D^M_{\mu }(\omega ) \ .
\end{eqnarray} 
The  result shown in Fig.\ 2 reveals that the impurity state is mainly
composed of
the \textit{M}-site states and that the change in the impurity band is mainly
ascribed to the change in $D^M_{\mu }(\omega )$. 

Next, we investigate the manner of coupling between the carrier spin and
the localized spin.
In Fig.\ 3(a) we show
the result for the optical carrier spin polarization $P(\omega )$,
where $P(\omega )$ is defined by
\begin{eqnarray}
P(\omega )  = \frac{D_{\downarrow }(\omega )-D_{\uparrow }(\omega )}
{D_{\downarrow }(\omega )+D_{\uparrow }(\omega )} \ .
\end{eqnarray} 
The result strongly suggests that 
\begin{subequations}
\begin{eqnarray}
P(\omega ) & \approx & +\frac{\langle S_z \rangle_{\rm av}}{S} \qquad
 \text{in the lower impurity band,}  \\
P(\omega ) & \approx & -\frac{\langle S_z \rangle_{\rm av}}{S} \qquad
 \text{in the higher impurity band}.
\end{eqnarray} 
\label{PWW}
\end{subequations}
We can deduce the result of Eq.\ (\ref{PWW}) by assuming 
 that the carrier spin always couples antiparallel (parallel) to the
localized spin
in the lower (higher) impurity band. 
To confirm the above picture,
we calculate the spin-coupling strength $Q(\omega )$ defined 
by 
(see Appendix)
\begin{eqnarray}
\label{Qomega}
Q(\omega ) & \equiv & - \left. \frac{\langle \delta (\omega -H) \ \sigma
\cdot \mathbf{S} \rangle /S}
{\langle \delta (\omega -H) \rangle} \right| _{ M \textrm{-site}}
\ .
\end{eqnarray} 
Thus, $Q(\omega )$ corresponds to $-\langle \cos \theta \rangle$,
where $\theta $ is the {angle} between the carrier spin and localized
spin at an \textit{M} site.
The result for $Q(\omega )$ shown in Fig.\ 3(b) clearly indicates  that 
\begin{subequations}
\begin{eqnarray}
Q(\omega ) & \approx & \ \ 1.0 \qquad
 \text{in the lower impurity band,}  \\
Q(\omega ) & \approx & -1.0 \qquad
 \text{in the higher impurity band.}
\end{eqnarray} 
\end{subequations}
Above results show that in the lower magnetic impurity band
the carrier spin is almost completely antiparallel
 with the fluctuating localized spin 
at an  \textit{M} site. 

%
%
In Fig.\ 4, we extract the low energy part of the DOS. 
A magnetic impurity band forms around the impurity level
 and its DOS has similar shape to the model band.
The total number of states in the impurity band per site
is equal to \textit{x} 
irrespective of the value of $\langle S_z\rangle$.
When $\langle S_z\rangle=S$, 
all states in the impurity band 
are down-spin states,
whereas  
the impurity band is composed of the same number ($x/2$)  
of up- and down-spin states when $\langle S_z\rangle=0$. 
The impurity band  has a larger bandwidth 
in the ferromagnetic state than in the paramagnetic state.
 The change in the bandwidth is understood as a  result of the change in
the effective
 hopping integral between Mn sites. When the coupling between the
 localized spin and the carrier spin is strong, the effective hopping
 between site $i$ and $j$ is reduced in proportion to $\cos
 \theta_{ij}/2$ where $\theta_{ij}$ is  the angle between the localized
spins at  site $i$ and
$j$.\cite{Anderson55} 
 In a fully polarized ferromagnetic state $\cos \theta_{ij}/2 $ is
always unity while  
 $\langle \cos \theta_{ij}/2\rangle =2/3$  if the localized spins are
completely randomly oriented. 
 The result shown in Fig. 3 certifies that Anderson-Hasegawa theory
 is applicable for the magnetic impurity band in this case. 
 Change in the bandwidth makes the energy of the ferromagnetic state
 lower  than that of the paramagnetic state when \textit{n} is small. 
 The energy gain increases with  increase of \textit{n} and reaches a
maximum  at
$n \sim x/2$. and then  gradually decreases to  vanishes at
$n \sim x$ as will be shown later. 
 \subsection{The case of $IS=-0.4\Delta $ and $E_M=0$}
Here we discuss the case of the exchange energy $IS=-0.4\Delta$
and the band offset energy $E_M=0.0$.
In this case the acceptor level does not appear in the dilute limit 
since $|E_M \pm IS|<0.5 \Delta $. 
To our knowledge no magnetic impurity band has been reported 
 in $A^{\rm II}_{1-x}{\rm Mn}_xB^{\rm VI}$-type DMS's
\cite{Furdyna88}.
Therefore  the present model
with the parameters of $|E_M \pm IS|<0.5 \Delta $ may correspond to II-VI-DMS's.
Numerical results are presented in Figs. 5 $\sim$ 7.
In Fig.\ \ref{II-VIDOS} the spin-polarized DOS is shown for various
values of
 $\langle S_z \rangle $. 
As can be seen in Fig.\ 6(b), 
the carrier states at the \textit{M} -site exist in the whole range of
band energy. 
The results for $P(\omega )$ and $Q(\omega )$ shown in Fig.\
\ref{II-VIPQ} suggest 
that coupling between the carrier spin and the localized spin 
is not strong except at the band edges,
which is consistent with the weak $ \langle S_z \rangle $ dependence of
$xD^M(\omega )$ 
 shown in Fig.\ \ref{II-VILOCAL}(b). 
 In Fig.\  \ref{II-VIDOS} the Fermi level $\varepsilon _F$ 
for $n=0.05$ is indicated by an arrow.

\subsection{Property of ferromagnetism in $E_M=0$}
In order to study the nature of carrier-induced
ferromagnetism, we first calculate the Curie temperature $T_c$ 
in a very simple way.
We  estimate $T_c$ using the energy of the paramagnetic state $E(0)$ and
that of the completely
ferromagnetic state $E(S)$ as 
\begin{eqnarray}
k_B T_c & = & \frac{2}{3x} \ 
 [E(0)-E(S)]  \ 
\end{eqnarray}
 assuming $b=0$ in Eq.\ (\ref{Expansion}).
Note that the CPA condition (\ref{CPAcondition}) 
results in a cubic equation when $\langle S_z \rangle _{\rm av}=S$ 
and a quartic equation when $\langle S_z \rangle _{\rm av}=0$. \cite{taka99,taka01a,taka01b}

In Fig.\ \ref{MagnTcIS} the result for $T_c$ is presented
as a function of $n$ for various values of $IS/\Delta $.
We immediately notice that there are two different types of behavior in 
$T_c$  as a function of $n$
 depending on the size of $|IS|/\Delta $.
When $|IS|/\Delta$ is small ($|IS|/\Delta  \lesssim 0.3$),
the ferromagnetism occurs over a wide range of $n$.
The Curie temperature gradually increases with the increase in $n$
and reaches a broad maximum. 
Then it gently decreases and vanishes  at a critical value $n_c$.
The maximum $T_c$ stays at a  low value, and $n_c $
is much larger than $x (=0.05)$. 
On the other hand,
when $|IS|/\Delta$ is large ($|IS|/\Delta \gtrsim 0.7$),
the ferromagnetism occurs in a narrow range of $n\ (\lesssim x$).
The $T_c$ rises steeply and reaches a maximum
 at $n_x \approx x/2$, and then
it decreases rapidly.
The maximum $T_c$ is high  and $n_c$ is less than but nearly equal to $x$.
These two different features can be seen clearly 
in Fig. \ref{MagnTcPhase} as well,
where $n_c$ and the maximum $T_c$ are depicted as functions of $n$.
The carrier density $n_x$ at which $T_c$ reaches the maximum is
also shown.
Two different characteristic features were also recognized in the
 phase diagrams  obtained in an earlier study
 with Ising localized spins.\cite{Kayanuma02}
When $|IS|/\Delta \lesssim 0.3$, the maximum $T_c$ is
approximately proportional to $(IS/\Delta )^2$.
This suggests that in the range of $|IS|/\Delta \lesssim 0.3$
the perturbative treatment on $IS/\Delta $ is available
and the RKKY-like mechanism is expected to operate for a moderate
carrier density.
With further increase in $|IS|/\Delta $ 
the maximum $T_c$ rises rapidly ($0.3 \lesssim |IS|/\Delta  \lesssim
0.7$), 
and then tends to saturate.
For $|IS|/\Delta  \gtrsim 0.7$,
ferromagnetism is induced only when $n \lesssim x$
and the maximum $T_c$ realizes at $n_x \cong x/2$.
The case with $IS = -\Delta$ studied in \S 3.2 belongs to this region 
and therefore the mechanism for ferromagnetism is understood from the argument above. 
As was shown in Fig. 4
the width of the magnetic impurity band increases
with increase of magnetization according to Anderson-Hasegawa mechanism.
It leads to the energy gain in ferromagnetic state
if the Fermi level lies in the impurity band.  
This is essentially the same mechanism with that of the double-exchange
 interaction  which causes ferromagnetism in mixed valency transition metal
 oxides such as (LaCa)MnO$_3$.\cite{Zener51,Anderson55} 
Therefore we may say that  the 
`double-exchange (DE)-like' mechanism for ferromagnetism 
is operative in the impurity band when $|IS|/\Delta \gtrsim 0.7$. 
Here we point out that the magnetic impurity levels appear when
$|IS|/\Delta >0.5$ if $E_M=0$.
Therefore we conclude that the DE-like mechanism in a magnetic impurity band
becomes dominant when $|IS|/\Delta  \gtrsim 0.7$.
In II-VI DMS's, the
 magnetic impurity level does not appear
as illustrated in Fig.\ \ref{II-VIDOS},
in which the parameters, 
$|IS|/\Delta =0.4$ and $E_M=0$, were employed.
Therefore, we may conclude
 that the DE-like mechanism is not relevant to the
 ferromagnetism in II-VI DMS's. 

To confirm above picture, we try to estimate $T_c$ in the 
strong exchange interaction limit 
through a simple argument.
In this limit 
the impurity band 
has the same shape of the model density of states $\rho (\omega )$
defined by Eq.\ (\ref{rho}) around
the impurity level $E_a$.  
In the completely ferromagnetic case 
 the DOS of the impurity band has a finite value
at the energies of $E_a- \sqrt{x} \Delta \leq \omega \leq E_a+ \sqrt{x}
\Delta $ 
and is given by
\begin{eqnarray}
D_F (\omega ) & = & \sqrt{x} \times \rho \left( \frac{\omega-E_a
}{\sqrt{x}} \right) \ .
\label{FerroDOST}
\end{eqnarray} 
In the paramagnetic case 
 the DOS of the impurity band has a finite value
at the energies of $E_a- \sqrt{x/2} \Delta \leq \omega \leq E_a+
\sqrt{x/2} \Delta $ 
and is given by
\begin{eqnarray}
D_P (\omega ) & = & 2 \times \sqrt{x/2} \times \rho \left(
\frac{\omega-E_a }{\sqrt{x/2}} \right) \ ,
\label{ParaDOST}
\end{eqnarray} 
where the factor 2 is due to up- and down-directions of the carrier spin.
The result for $T_c$ based on the assumption is inserted in Fig.
\ref{MagnTcIS} as a 'LIMIT'.
The maximum $T_c$ is estimated to be 
\begin{eqnarray}
k_B T_c & = & \frac{2(2-\sqrt{2})}{9\pi } \sqrt{x} \Delta \ ,
\label{LIMIT}
\end{eqnarray} 
at $n = x/2$. The value of the maximum $T_c\ (=0.0093\Delta )$ for
$x=0.05$ is
pointed by an arrow in Fig. \ref{MagnTcPhase}.

%
It is worth noting that the Zener double-exchange mechanism
for ferromagnetism is usually understood to be effective 
 only when the exchange energy is larger than the width of the carrier
band 
(or $|IS| \gtrsim  2\Delta $) in  the case where magnetic ions sit on each
site. 
In the present case the exchange energy is not larger than the width of
 the model band.
Nevertheless the `DE-like' mechanism functions 
because the magnetic impurity bandwidth is smaller than the exchange energy
(or $|IS| \gtrsim \sqrt{2x }\Delta $).

\subsection{Effect of the attractive potential }
\begin{figure}[p]
\begin{center}
\includegraphics[width=7.cm,clip]{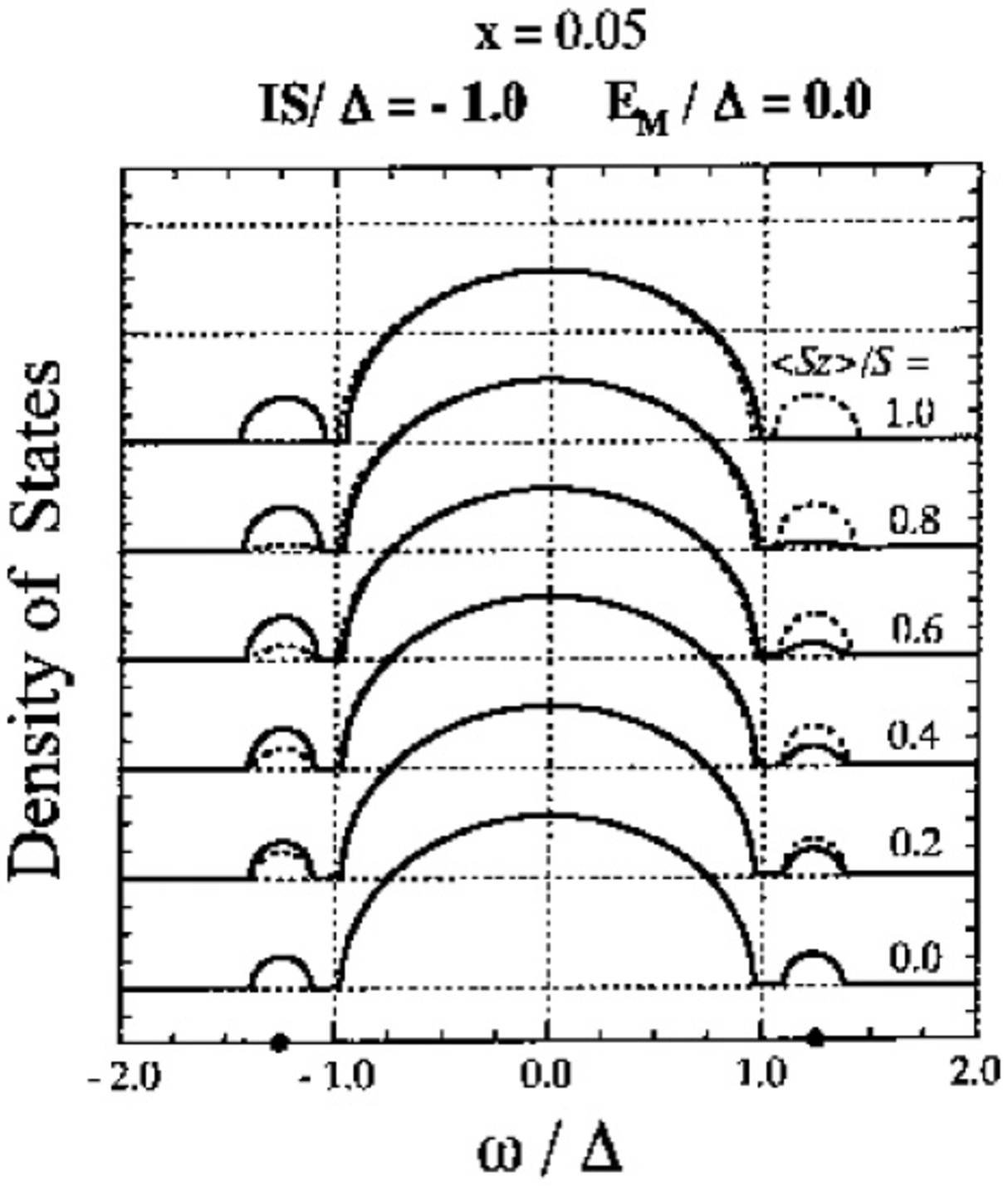}
\caption{\label{strongDOS}
 The result for the DMS with $IS=-\Delta $ and $E_M=0$.
The density of states (DOS) as a function of $\omega /\Delta $ 
for various values of $\langle S_z \rangle  /S$.
Solid line is for down-spin carrier, and dotted line is for up-spin carrier.
The impurity levels $E_a = \pm 1.25 \Delta$ are indicated by dots
 on the abscissa.
}
\end{center}
\end{figure}

We study here the role of the attractive potential in order to elucidate
 the origin of
 the carrier-induced ferromagnetism
in III-V-based DMS's.
Figures \ \ref{deepDOS} $\sim$  \ref{deepPQ}
show the results for the case where $IS=-0.4 \Delta $ and 
$E_M=-0.6 \Delta $.
Although the strength of the exchange coupling is same
 as that in \S 3.3,
 an impurity band appears 
due to an attractive nonmagnetic local potential $E_M=-0.6 \Delta $.
Furthermore, we find a strong similarity between  the low energy part 
of  DOS in Fig.\ 1 ($IS=-\Delta $ and $E_M=0$) 
and  that  in Fig.\ \ref{deepDOS} ($IS=-0.4\Delta $ and $E_M=-0.6\Delta $).
The similarity is due to the fact that the impurity level has the same
 energy,
 $E_a =-1.25\Delta $, which is determined by the effective attractive
potential $E_M+IS$. 
When $\langle S_z \rangle=0 $
a magnetic impurity band with  $x/2$ down- and up-spin states forms
 around the impurity level in both cases. 
Thus the impurity bands in both cases  have a strong similarity
although they do not completely agree with each other. 
When $\langle S_z \rangle=S $, 
 the down-spin bands
agree completely with each other in  both cases 
 because the DOS's were determined only by the energy of antiparallel
coupling 
 $E_M +IS$. 
 The up-spin impurity bands are completely suppressed in both cases
and the states merge into the host band.
Thus, for  up-spin as well, 
we find a similarity in the low energy part of the DOS's   
shown in Figs. \ref{strongDOS} and \ref{deepDOS}.
 From the similarity in the low energy part of the DOS, we may expect
that 
ferromagnetism occurs through the same mechanism in both cases. 
In Fig. \ref{MagTcIS} the effect of the nonmagnetic potential $E_M$ on $T_c$
is presented for  $IS$ fixed to be $-0.4 \Delta $.  The impurity level
 appears for $E_M<-0.1 \Delta$ in this case. 
When $E_M \gtrsim 0.0$ the $T_c$ stays low  and $n_c $ is much larger
than  $x$, while for $E_M \lesssim -0.2 \Delta $ 
 high $T_c$ is realized and $n_c$ is less than  $x$.
 In the latter region the DE-like mechanism becomes in effect. 
Figure \ref{MagnTcEM} shows how the maximum $T_c$ as well as 
$n_c$  varies with  $E_M$ and $IS$. 
The result reveals that the attractive nonmagnetic potential $E_M$
extends the region where the DE-like mechanism is effective  and 
enhances maximum $T_c$.
The criterion for the DE-like mechanism to operate
is roughly estimated to be $IS+0.4E_M \lesssim -0.6 \Delta $.


%
\section{ Summary and Discussions}
 In the present study
we applied the dynamical CPA to a simple model in order 
to investigate the carrier states of DMS's in a systematic way.
The present results reveal
that a  magnetic impurity band forms  separate from  the host band
when the effective attractive potential is strong enough.
In such a case  the spin of a carrier in the impurity
band couples very  strongly with the localized spin  at  a magnetic
impurity site. Then   
the hopping of the carriers  among the magnetic sites
aligns the localized spins parallel
through the DE-like mechanism.
The effective attractive potential is determined by the sum of the
 non-magnetic attractive potential and the exchange interaction. 
Therefore, the non-magnetic  potential substantially enhances the effect
of the exchange interaction. This mechanism
works even in the case where 
 the impurity band is not completely separate from the host band.
This is the case in III-V-based DMS's, where the exchange interaction
alone is not strong enough to make a pronounced impurity band. 
On the other hand, the  absence of non-magnetic attractive potential in 
II-VI-based DMS's suppresses the magnetic impurity band and therefore the
appearance of strong ferromagnetism. 
 
Here we compare  the Zener double-exchange (DE) mechanism 
 and the present DE-like mechanism that occurs in the magnetic impurity
band. 
Zener originally proposed the DE interaction for (La,Ca)MnO$_3$ where
 3\textit{d}-holes
hop  among the magnetic ions sitting on the regular lattice sites
 \cite{Jonker50}. 
Therefore it might be usually understood that DE mechanism is only
 relevant to hopping of the 3\textit{d} holes in (Ga,Mn)As. 
The DE mechanism for ferromagnetism is, in fact,
 works quite generally in cases where the kinetic  energy (or the
bandwidth) of
 a carrier is not greater than the exchange energy between a carrier
 spin and
 a localized spin. It originates from the effective reduction of the
 hopping amplitude
 between sites with localized spins which  are not parallel with each
 other. The reduction is in proportion to $\cos \theta/2$ where
 $\theta$ is the relative angle between the localized
spins\cite{Anderson55}. 
Only condition required for this reduction is a very strong exchange
 coupling between carrier spins
 and localized spins. If it is satisfied, 
 carriers may have any character and the localized spins can
 be arranged randomly. 
Even localized spins are not necessary for this mechanism if strong
 Hund couplings are  working between carrier spins  on the same
site\cite{Momoi98,Sakamoto02}. 

In the case of III-V based DMS's, carriers are considered to have 
4\textit{p} character
 \cite{Oka98,Ishiwata02} and therefore the bandwidth ($\sim$4 eV) is not
 larger than the exchange energy ($IS= 0.6\sim 0.8$ eV).  
However the attractive interaction between the carrier and a Mn$^{2+}$
 ion helps for the ion to bind the carrier,
 and in consequence the carrier
 moves in a narrow 
impurity band. The effective  
bandwidth for a carrier is that of the impurity band 
and it is smaller than the exchange energy.
The DE mechanism 
becomes in effect even if the impurity band is not completely
separated from the host band.
 We call this mechanism ``DE-like''
 mechanism in order to avoid
 confusion with the argument assuming $d$ holes \cite{Akai98}.
We note here that no Mn$^{3+}$ ($d^4$ configuration) states have been
 experimentally detected in (Ga,Mn)As
\cite{Linnarsson97,Oka98,Szczytko99,Szczytko99b,Ohldag00}. 
All these experimental observations rather suggests the picture that
 the fixed valence state  Mn$^{2+}$ ($S=5/2$) is realized in (Ga,Mn)As.
The absence of the
 attractive potential suppresses the appearance of a magnetic impurity
 band and therefore the DE-like mechanism does not work 
 in II-VI-based DMS's.  

The results for the $n$ dependence of the energy gain and $T_c$,
 shown in Fig.\ \ref{GaMnAsTc}
 are well explained on the basis of the DE-like mechanism.
Recently Hayashi \textit{et al.} reported that heat treatments
(annealing) could
vary the hole concentration $n_h$ while holding  $x$
constant \cite{Hayashi01}. 
They reported in a sample
with a nominal Mn content of 5 \% that 
 $T_c$ increased  from  $\sim$80K to $\sim$95K and then decreased again
   to $\sim$80K 
when  $n_h$ was increased from 6.1 to 6.8 and then 7.2 $ \times
10^{19}$ cm$^{-3}$. 
The carrier concentration in these cases are 2.7, 3.1  and
3.3\%, respectively. The result implies that   $T_c$ takes its
maximum at $n_h \sim 0.6x$ in the sample, while the present theory
 predicts a maximum at $\sim 0.3x$ as is  shown in Fig. 20(b).  
Study of a more realistic model taking 
into other effects 
and/or more detailed experimental investigations are necessary  to
clarify this discrepancy.

Although we assumed the \textit{p}-holes with a wide bandwidth for carriers,
the present result is in contrast with the RKKY argument which assumes 
the free-carrier picture
\cite{Matsukura98}. 
Infrared optical absorption measurements  
have recently revealed that the hole wave functions are nearly localized 
and the transport should be more or less hopping-like 
even in the most metallic Ga$_{1-x}$Mn$_x$As \cite{Hirakawa02,Katsumoto01}.
The nearly bound-hole picture
is consistent with the present result that the \textit{p} hole in the
band tail  
prefers to stay on Mn sites
[see Fig.\ \ref{GaMnAsPQ}(c)].

Although the exchange interaction between \textit{p} hole and 
\textit{d} spin is experimentally established  to be \textit{antiferro}magnetic
\cite{Oka98,Ando98,Szczytko99},
the  exchange interaction
was reported to be  \textit{ferro}magnetic in an early study on the
polarized
 magnetoreflection\cite{Szcytko96}.
In our result shown in   Fig.\ \ref{strongFIG},  
the bottom of the down-spin magnetic impurity band shifts to the lower
energy side
 with increase in $\langle S_z \rangle$,
whereas that of the host band shifts to the higher energy side. 
It is also worth noting that $Q(\omega ) \approx -1.0$
at the bottom of the host band (see  Fig.\ \ref{strongPQ} (b)). 
These indicate that the  ferromagnetic spin coupling occurs
at the bottom of the host band, contrary to  the antiferromagnetic spin
coupling in the magnetic impurity band.
If we interpret that 
 the optically observed band edge 
is not that of the impurity band but that of the host band, the present
result may explain the experimental observation. 
A simple calculation based on the present approach consistently explains
 the apparent ferromagnetic exchange interaction experimentally
observed. 
Details will be published elsewhere \cite{Takahashi03}.  

Next, we briefly mention the quantum effect of the localized spin
that is ignored in the present work.
When the calculations are done for the quantum spins with finite $S$,
the total number of states in a impurity band is not $x$ but
$2x(S+1)/(2S+1)$ for parallel coupling 
and $2xS/(2S+1)$ for antiparallel coupling, respectively.
The result is valid
 as long as the band is almost empty.
The many body effect, however, may change the situation \cite{Edwards99}.
Therefore, although the present result suggests that 
$T_c$ might  be multiplied by a factor $\sqrt{2(S+1)/(2S+1)}$ 
or  $\sqrt{2S/(2S+1)}$  in the strong coupling limit,
the effect of quantum spin is still unclear at the present stage.

The Curie temperature  obtained by using  the  model 
with  randomly distributed Ising spins \cite{Kayanuma02}
is about three times larger than the present result.  
The reason  may be due to the fact  that the spin-flip effect
 is not taken into account  in their theory. 

Throughout this work, the carriers were assumed to be degenerate.
Recently the dynamical mean-field theory \cite{Georges96,Furukawa99} was applied
to the present model.
The result for $T_c$ is almost the same as the present result,
which suggests that the assumption of the degenerate carriers is reasonable.
The result will be published elsewhere together with the result of  the 
temperature dependence of magnetization \cite{Furukawa03}.

The present model  has succeeded  in explaining  the fundamental 
mechanism of the carrier-induced ferromagnetism in III-V based
DMS's.
There are, however, many features which exist in the real DMS's but  are
not taken into account.  Those are multi-band effects, the band anisotropy,
 the spin-orbit coupling, \textit{p-d} hybridization,
 the long-range 
character of the  Coulomb potential, the direct antiferromagnetic
exchange interaction
between Mn ions and so on. 
These issues remain for future study.



\clearpage

\begin{figure}[p]
\begin{center}
\includegraphics[width=6.cm,clip]{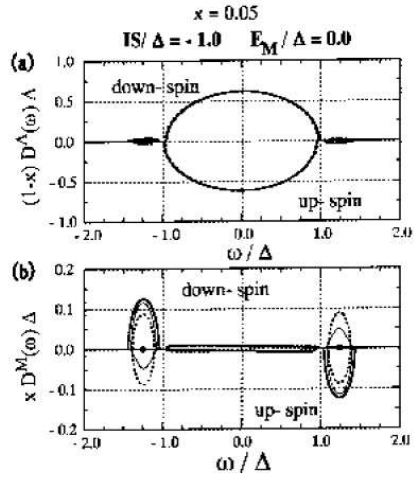}
\caption{\label{strongLOCAL}
 The results for the DMS with $IS=-\Delta $ and $E_M=0$.
(a) \textit{A}-site component of the DOS,
 $(1-x) D^A_{\downarrow }(\omega )\Delta $ and  $-(1-x) D^A_{\uparrow
}(\omega )\Delta $.
(b) \textit{M}-site component of the DOS,
 $x D^M_{\downarrow }(\omega )\Delta $ and  $-x D^M_{\uparrow }(\omega
)\Delta $.
The thick, thin and dotted lines represent the cases of 
 $\langle S_z \rangle /S=1.0$, 
$0.5$ and 0.0, respectively.  
Note the difference in the scale of vertical axes between (a) and (b). 
}
\end{center}
\end{figure}

\begin{figure}[p]
\begin{center}
\includegraphics[width=6.cm,clip]{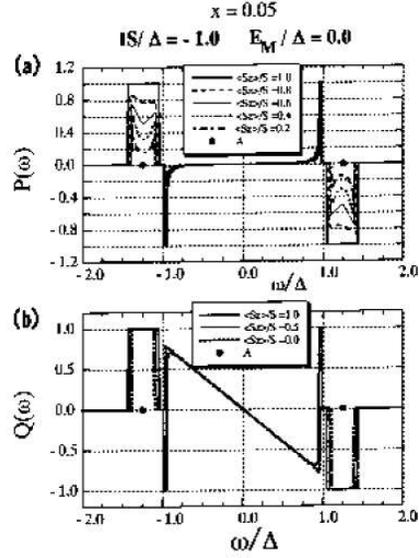}
\caption{\label{strongPQ}
 The result for the DMS with $IS=-\Delta $ and $E_M=0$.
(a) Optical carrier-spin polarization $P(\omega )$
for $\langle S_z \rangle /S=$ 1.0,0.8,0.6,0.4,0.2 and 0.0.  
(b) Spin-coupling strength $Q(\omega )$.
The thick, thin and dotted lines represent the cases of 
 $\langle S_z \rangle /S=1.0$, 
$0.5$ and 0.0, respectively.  
The impurity levels $E_a = \pm 1.25 \Delta$ 
are indicated by dots on the abscissa.
}
\end{center}
\end{figure}

\begin{figure}[p]
\begin{center}
\includegraphics[width=7.cm,clip]{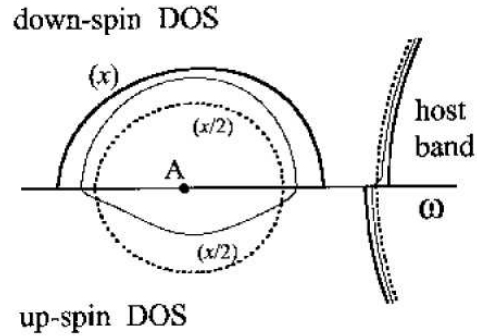}
\caption{\label{strongFIG}
 The DOS of the impurity band in the case of $IS/\Delta = -1.0$ 
and $E_M = 0$.
The thick, thin and dotted lines represent the cases of 
 $\langle S_z \rangle /S=1.0$, 
$0.5$ and 0.0, respectively.  
The dot A indicates the impurity level for $x\rightarrow 0$.}
\end{center}
\end{figure}

\begin{figure}[p]
\begin{center}
\includegraphics[width=7.cm,clip]{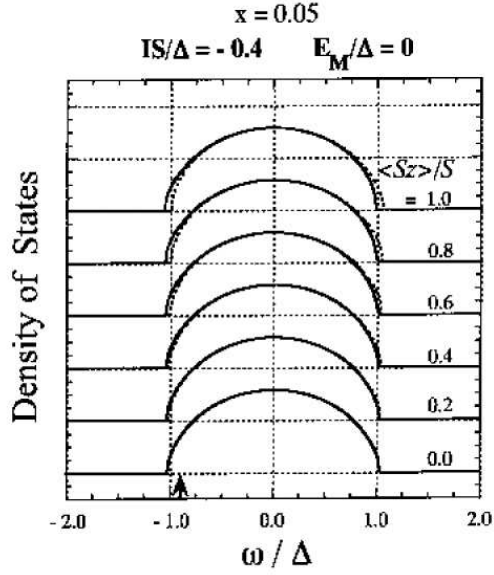}
\caption{\label{II-VIDOS}
 The result for $IS=-0.4\Delta $ and $E_M=0$.
 The DOS as a function of $\omega /\Delta $
for various values of $\langle S_z \rangle /S$.
Solid line is for down-spin carrier, and dotted line is
for up-spin carrier.
The arrow points to the Fermi level $\varepsilon_F/\Delta$ for $ n = x\
(=0.05) $.
}
\end{center}
\end{figure}

\begin{figure}[p]
\begin{center}
\includegraphics[width=6.cm,clip]{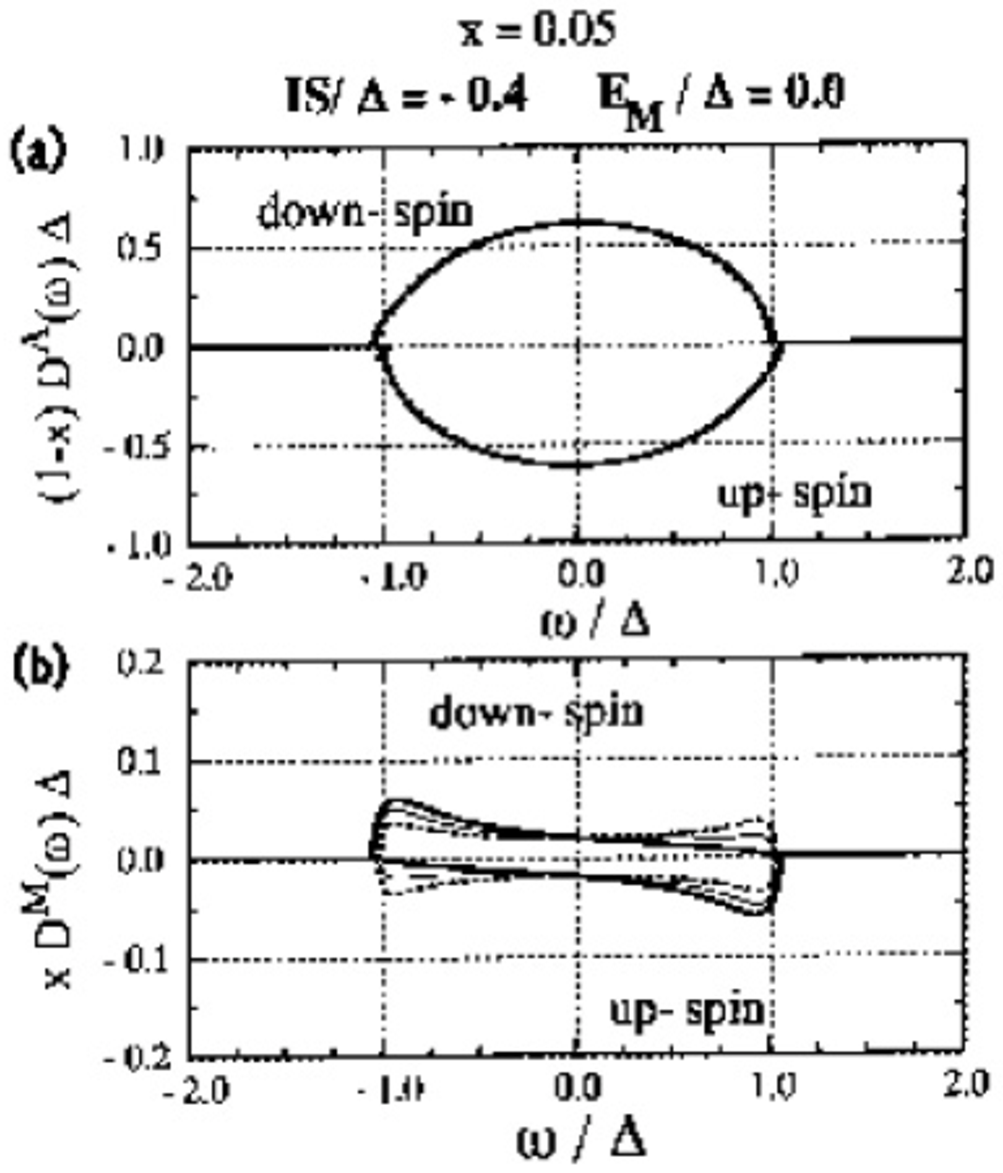}
\caption{\label{II-VILOCAL}
 The result for the DMS with $IS=-0.4\Delta $ and $E_M=0$.
(a) \textit{A}-site component of the DOS, 
$(1-x) D^A_{\downarrow }(\omega )\Delta $ and  $-(1-x) D^A_{\uparrow
}(\omega )\Delta $.
(b) \textit{M}-site component of the DOS,
 $x D^M_{\downarrow }(\omega )\Delta $ and  $-x D^M_{\uparrow }(\omega
)\Delta $.
The thick, thin and dotted lines represent the cases of 
 $\langle S_z \rangle /S=1.0$, 
$0.5$ and 0.0, respectively.  
}
\end{center}
\end{figure}
\clearpage
\begin{figure}[p]
\begin{center}
\includegraphics[width=6.cm,clip]{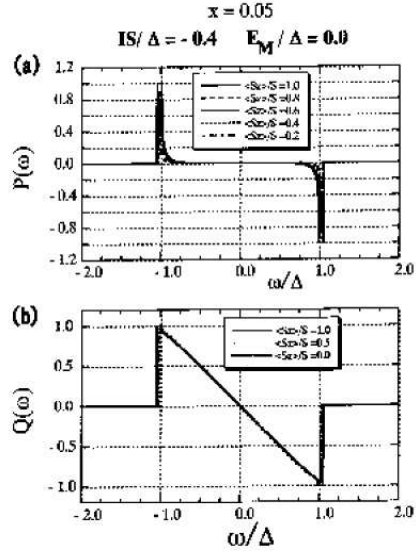}
\caption{\label{II-VIPQ}
 The result for the DMS with $IS=-0.4\Delta $ and $E_M=0$.
(a) optical carrier-spin polarization $P(\omega )$,  
(b) spin-coupling strength $Q(\omega )$.
}
\end{center}
\end{figure}

\begin{figure}[p]
\begin{center}
\includegraphics[width=7.cm,clip]{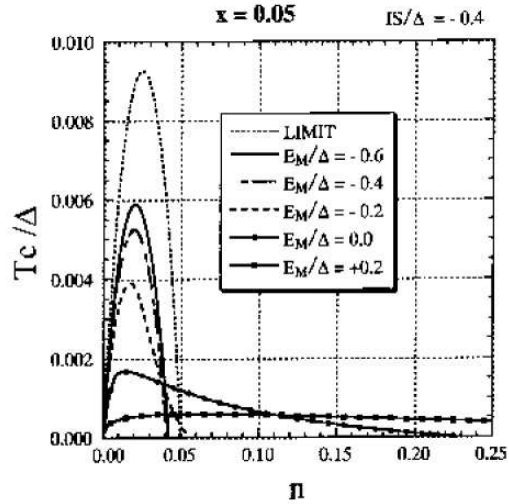}
\caption{\label{MagnTcIS}
 The result for Curie temperature $T_c/\Delta $ as
a function of carrier density $n$ for various values of 
$IS/\Delta $ with $x=0.05$ and $E_M=0$.
The result based on the assumption that an impurity band 
is similar shape of the model band is 
drawn as 'LIMIT' (see text).}
\end{center}
\end{figure}

\begin{figure}[p]
\begin{center}
\includegraphics[width=7.cm,clip]{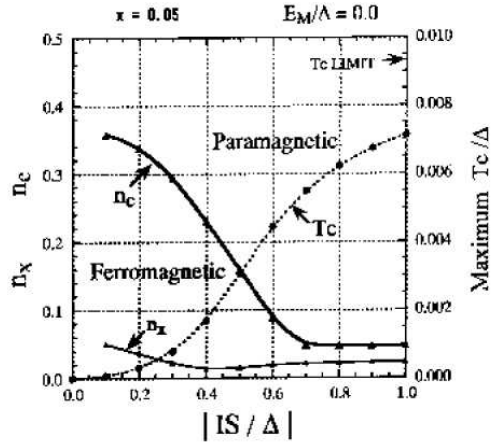}
\caption{\label{MagnTcPhase}
 Phase diagram for $E_M =0$ and $x=0.05$.
The critical value $n_c$ (solid line; left scale)
 and the maximum $T_c$ (dotted line; right scale)
are presented as a function of $|IS|/\Delta$.
The carrier density $n_x$ at which $T_c$ reaches the maximum is included
(solid line; left scale).}
\end{center}
\end{figure}

\begin{figure}[p]
\begin{center}
\includegraphics[width=7.cm,clip]{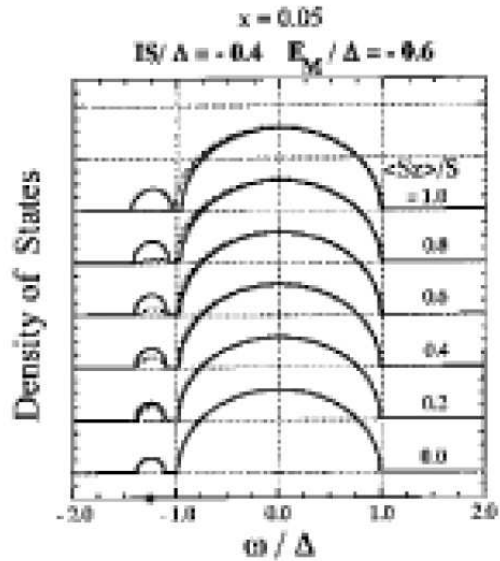}

\caption{\label{deepDOS}
  The result for the DMS with  $IS=-0.4\Delta $ and $E_M=-0.6\Delta $.
The DOS is plotted as a function of $\omega /\Delta $
for various values of $\langle S_z \rangle  /S$.
Solid line is for down-spin carrier, and dotted line is
for up-spin carrier.
The impurity level $E_a = -1.25 \Delta$ 
is indicated by the dot on the abscissa.
}
\end{center}
\end{figure}

\begin{figure}[p]
\begin{center}
\includegraphics[width=6.cm,clip]{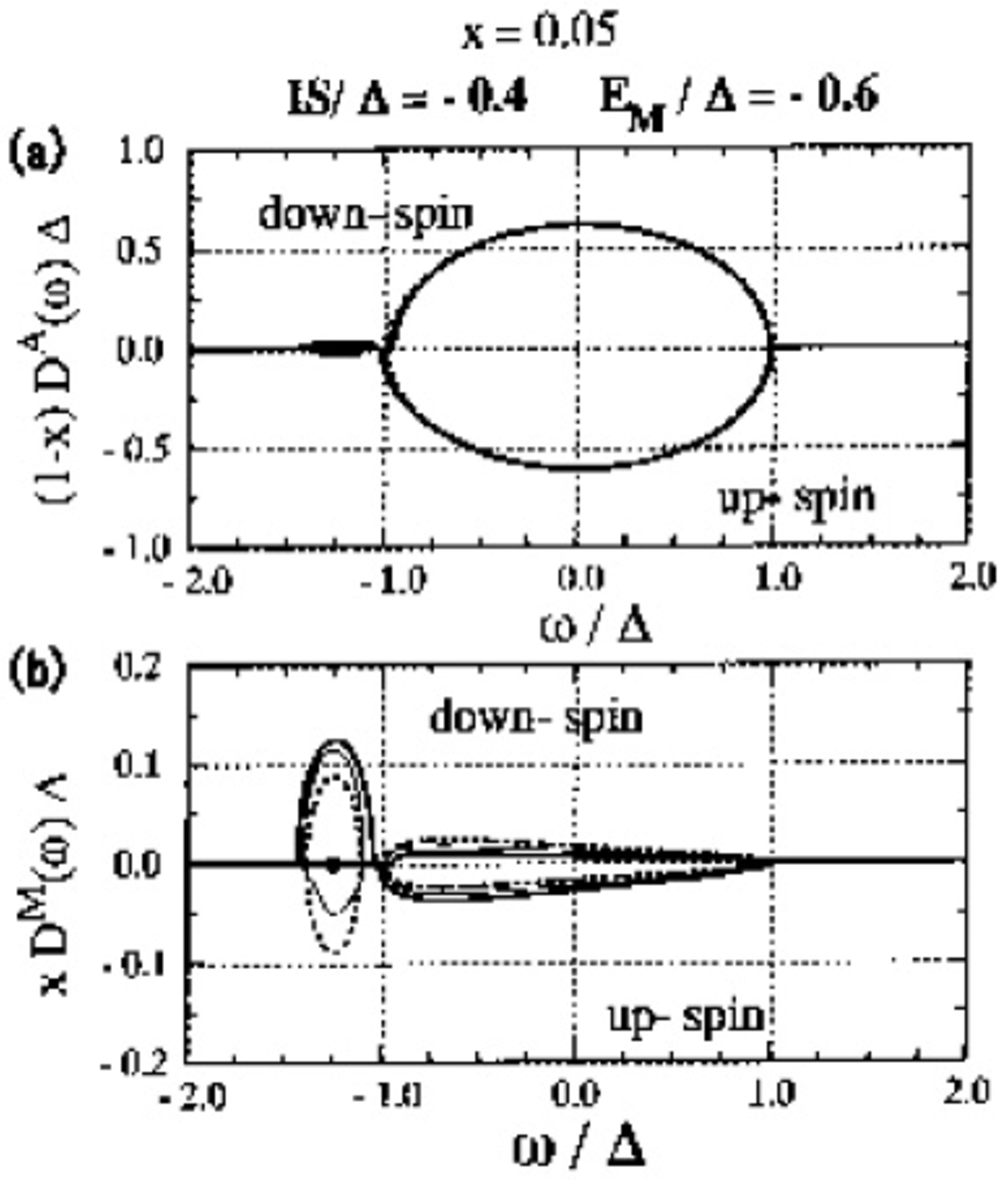}
\caption{\label{deepLOCAL}
 The result for the DMS with $IS=-0.4\Delta $ and $E_M=-0.6\Delta $:
(a) \textit{A}-site component of the DOS,
 $(1-x)D^A_{\downarrow }(\omega )\Delta $ and  $-(1-x)D^A_{\uparrow
}(\omega )\Delta $,
(b) \textit{M}-site component of the DOS, 
$x D^M_{\downarrow }(\omega )\Delta $ and  $-x D^M_{\uparrow }(\omega
)\Delta $.
The thick, thin and dotted lines represent the cases of 
 $\langle S_z \rangle /S=1.0$, 
$0.5$ and 0.0, respectively.  
}
\end{center}
\end{figure}

\begin{figure}[p]
\begin{center}
\includegraphics[width=6.cm,clip]{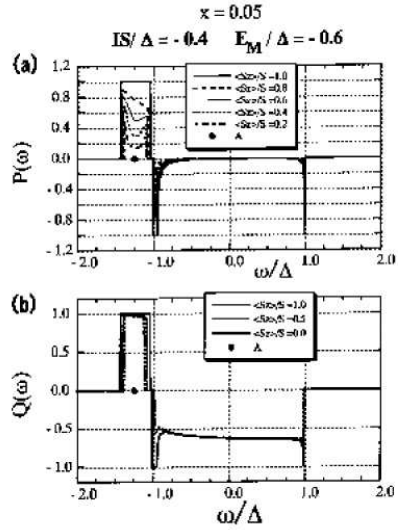}
\caption{\label{deepPQ}
 The result for the DMS with  $IS=-0.4\Delta $ and $E_M=-0.6\Delta $:
(a) optical carrier-spin polarization $P(\omega )$, 
(b) spin-coupling strength $Q(\omega )$.
The impurity level $E_a = -1.25 \Delta$ 
is indicated by a dot on the abscissa.
}
\end{center}
\end{figure}
\clearpage

\begin{figure}[p]
\begin{center}
\includegraphics[width=7.cm,clip]{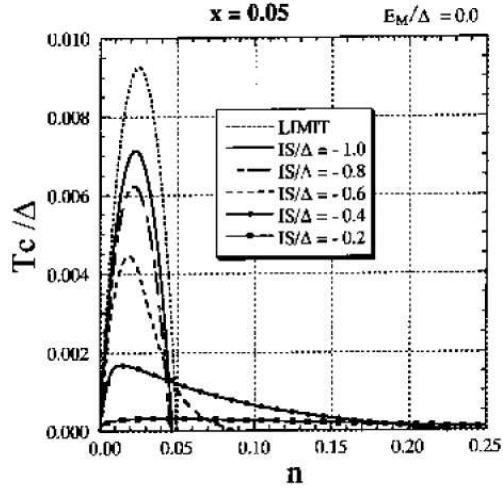}
\caption{\label{MagTcIS}
The result for $T_c/\Delta $ as
a function of $n$ for various values of 
$E_M/\Delta $ with $x=0.05$ and $IS= -0.4 \Delta $.
The result of 'LIMIT' is included (see text).}
\end{center}
\end{figure}

\begin{figure}[p]
\begin{center}
\includegraphics[width=7.cm,clip]{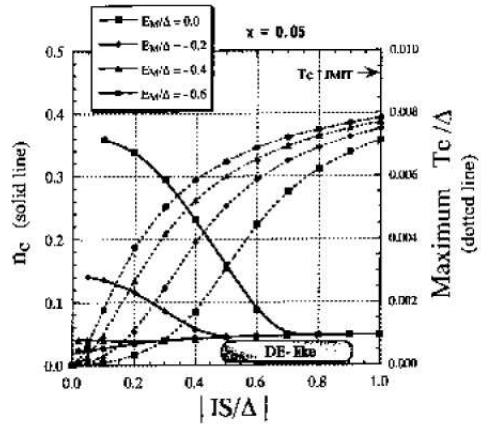}
\caption{\label{MagnTcEM}
 Phase diagram for various values of $E_M/\Delta$.
The critical value $n_c$ (solid line; left scale)
 and the maximum $T_c$ (dotted line; right scale)
are presented as a function of $|IS|/\Delta $
for various values of $E_M /\Delta$.} 
\end{center}
\end{figure}
\end{document}